\documentclass[aps,prl,twocolumn,showpacs,superscriptaddress]{revtex4}
\usepackage{hyperref}
\usepackage{graphicx}
\usepackage{amsmath}
\usepackage{amssymb}
\usepackage{lscape}
\usepackage{braket}
\usepackage{xcolor}

\input{epsf}

\begin{document}

\title{Density Oscillations Induced by Individual Ultracold Two-Body Collisions}

\author{Q. Guan}
\affiliation{Homer L. Dodge 
Department of Physics and Astronomy, The University of Oklahoma, 
440 West Brooks Street,
Norman, Oklahoma
73019, USA}

\author{V. Klinkhamer}
\author{R. Klemt}
\author{J. H. Becher}
\author{A. Bergschneider}
\altaffiliation[Current address: ]{Institute of Quantum Electronics, ETH Zurich,
CH-8093 Zurich, Switzerland}
\author{P. M. Preiss}
\author{S. Jochim}
\affiliation{Physics Institute, Heidelberg University,
Im Neuenheimer Feld 226, 69120 Heidelberg, Germany}

\author{D. Blume}
\affiliation{Homer L. Dodge 
Department of Physics and Astronomy, The University of Oklahoma, 
440 West Brooks Street,
Norman, Oklahoma
73019, USA}

\date{\today}

\begin{abstract}
Access to single particle momenta provides new means of studying the dynamics of few interacting particles. In a joint theoretical and experimental effort, we observe and analyze the effects of a finite number of ultracold two-body collisions on the relative and single-particle densities by quenching two ultracold atoms with initial narrow wave packet into a wide trap with inverted aspect ratio. The experimentally observed spatial oscillations of the relative density are reproduced by a parameter-free zero-range theory and interpreted in terms of cross-dimensional flux. We theoretically study the long time dynamics and find that the system does not approach its thermodynamic limit. The set-up can be viewed as an advanced particle-collider that allows one to watch the collision process itself. 

\end{abstract}

\maketitle

The one-dimensional harmonic oscillator is discussed
in many 
text books, 
from introductory classical 
and quantum mechanics to quantum optics and field theory~\cite{ho_book}.
The physics of the one-dimensional
harmonic oscillator is simple:
Its classical orbits are sinusoidal and periodic and the
quantum propagator has a compact analytical expression.
Moreover, the harmonic oscillator allows one to gain intuition
for the dynamics of multi-dimensional systems.

This work studies, both experimentally and theoretically,
the quench dynamics of an {\em{anisotropic}} three-dimensional
harmonic oscillator in which the three degrees of freedom are coupled by a point
scatterer of varying strength that is located at the origin.
Since the point scatterer has a measure of zero, 
the classical trajectories 
are not influenced by the point scatterer~\cite{Seba}.
However, the situation changes
drastically
when one enters the quantum regime since the point scatterer can
simultaneously partially reflect
and partially transmit the wave packet, or even reflect the wave packet in its entirety
~\cite{Seba, Seba_2, Seba_3, Olshanii_1, Olshanii_2, Olshanii_3, Qu, John_1, John_2, John_3, artem, blume, peter, busch_dynamics}.

The quench dynamics of one-dimensional quantum systems has been investigated
extensively at the microscopic level~\cite{1d_exp_1, 1d_exp_2, 1d_exp_3, Newton_Cradle_1, Newton_Cradle_2, Newton_Cradle_3, Schmiedmayer_1d, 1d_theory_1, 1d_theory_2, John_2, John_3, artem, blume, peter, busch_dynamics}. 
Examples include the realization of quantum Newton's cradle~\cite{Newton_Cradle_1, Newton_Cradle_2, Newton_Cradle_3} and the 
observation of quantum revivals in a system containing 
around
$1000$ atoms, addressing questions related to equilibration, thermalization, and 
their connections to integrability of one-dimensional systems~\cite{Schmiedmayer_1d}.
The quench dynamics of three-dimensional systems is expected to differ from that 
of one-dimensional systems in important ways.  
This letter explores these differences by studying the quench dynamics of an anisotropic harmonic oscillator, 
including the weakly-attractive and repulsive regimes, 
where the system behavior is quite intuitive, 
and the strongly-interacting regime,
where the $s$-wave scattering length is the largest length scale in the problem and intuition tends to fail. 

\begin{figure*}
\vspace{0cm}
\includegraphics{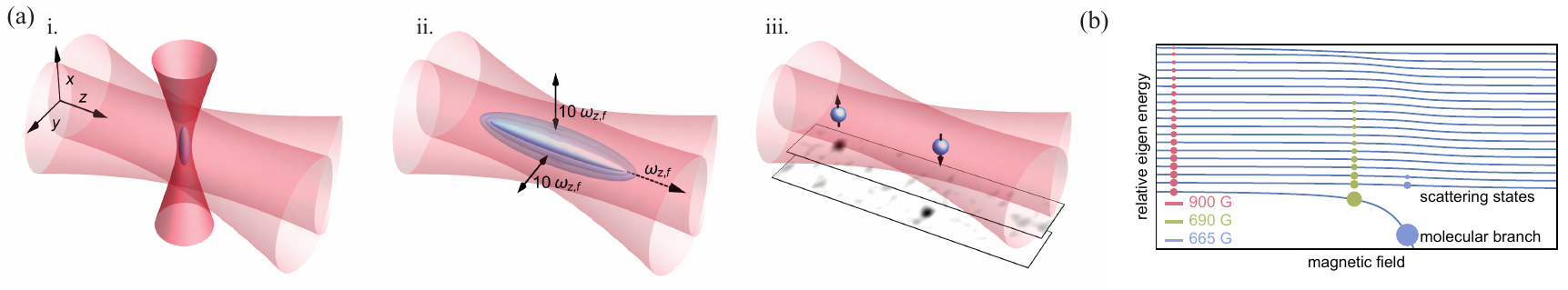}             
\vspace{-0cm}
\caption{(color online)
Schematic of system and characterization
of initial state. 
(ai) We 
prepare the wavepacket of two interacting $^6$Li atoms 
(blue cloud) in the ground state of a tight optical tweezer trap (dark red). 
We additionaly ramp on a weak crossed beam optical dipole trap (light red, not to scale). 
(aii) At $t=0$ we switch off the tweezer trap to quench the trap geometry 
so that the cloud quickly oscillates in the $x$- and $y$-directions while slowly expanding in the $z$-direction. 
(aiii) After a variable expansion time we record the $z$-positions of both atoms via fluorescence imaging.
(b) Schematic energy spectrum of the Hamiltonian $H_{\text{rel},f}$ after the quench, 
showing a single ``molecular branch'' and a nearly harmonic spectrum of ``scattering states''. 
Different magnetic fields
lead to very different projections of the initial state onto the eigenstates of $H_{\text{rel},f}$. 
In the weakly-interacting regime at large magnetic fields, 
many scattering states are populated with roughly equal probability. 
At lower magnetic fields, the projection onto the single molecular state increases, 
eventually dominating the two-body dynamics. 
For the 
experimental parameters, hundreds of levels contribute to the dynamics.
}
\label{fig1}
\end{figure*}

We realize
the three-dimensional anisotropic harmonic oscillator
with point scatterer experimentally by optically trapping two ultracold atoms 
[Fig.~\ref{fig1}(a)], which interact
via a short-range van der Waals potential with tunable scattering length. 
The dynamics are initiated by a quench of the trap geometry.
The system provides a versatile platform
for studying few-body dynamics in  
a regime where a small and predictable number of collisions occur. 
Since the optical trap is nearly perfectly harmonic,
the center-of-mass motion,
which is not affected by the interactions,
decouples from the relative motion.
Thus, we focus on 
(i) the dynamics in the relative
degrees of freedom and 
(ii) the impact of this motion
on the single-particle density. 
Excellent agreement with our parameter-free
theory predictions is found.
In the strongly-interacting regime, the resulting density profiles in the relative,
low-energy
$z$-coordinate 
display time-dependent oscillatory or fringe pattern, which we interpret as
signatures
of cross-dimensional dynamics.
The single-particle density profiles, in contrast,
are smooth except for 
very short time periods during which the two particles are close to each other.
This illustrates that the scattering events
impact
the
single- and higher-order correlation functions differently. While the relative density varies appreciably with time, our calculations reveal an extremely slow approach to equilibrium, manifest in a failure to thermalize over thousands of cycles.

Experimentally, we prepare two $^6$Li atoms in two distinct hyperfine states denoted by $|1 \rangle$ and $|3 \rangle$~\cite{feshbach} 
in the motional ground state of a tightly focussed optical tweezer trap elongated along the $x$-direction 
[Fig.~\ref{fig1}(ai)]~\cite{exp1}. 
At time $t=0$, the system is quenched by instantaneously changing 
the trap geometry and aspect ratio 
[Fig.~\ref{fig1}(aii)]. 
We release the atoms into a much weaker dipole trap with inverted geometry,
whose weakest frequency is along the $z$-axis. 
Since the trap potentials are harmonic to a good approximation, 
the system before and after the quench is described in terms of the low-energy two-particle Hamiltonian $H_{\alpha}=H_{\text{rel},\alpha} + H_{\text{cm},\alpha}$,
where $\alpha=i,f$ denotes the geometry before ($i$) and after ($f$) the quench,
\begin{eqnarray}
H_{\text{rel},\alpha}= \frac{-\hbar^2}{2\mu} \vec{\nabla}^2_{\vec{r}} +
\frac{\mu}{2} (\omega_{x,\alpha}^2 x^2 + \omega_{y,\alpha}^2 y^2 + \omega_{z,\alpha}^2 z^2)
+ \nonumber \\
\frac{2 \pi \hbar^2 a_s}{\mu} \delta({\vec{r}}) \frac{\partial}{\partial r} r
\end{eqnarray}
and
\begin{eqnarray}
H_{\text{cm},\alpha}=
\frac{-\hbar^2}{2 M} \vec{\nabla}^2_{\vec{R}} +
\frac{M}{2}(\omega_{x,\alpha}^2 X^2 + \omega_{y,\alpha}^2 Y^2 + \omega_{z,\alpha}^2 Z^2)
.
\end{eqnarray}
Here, $\vec{r}=(x,y,z)^T$ and $\vec{R}=(X,Y,Z)^T$ are the relative
and center-of-mass position vectors, respectively, $\mu$ and $M$
the associated masses, and $a_s$ the three-dimensional $s$-wave scattering length characterizing the interaction strength.   

The experimentally measured trapping frequencies are $\omega_{x,i}:\omega_{y,i}:\omega_{z,i}=2\pi\times \left(6.4:31:30\right)$~kHz before and $\omega_{x,f}:\omega_{y,f}:\omega_{z,f}=2\pi\times \left(640:600:61.7\right)$~Hz (aspect ratio of $\omega_{x/y,f}/\omega_{z,f}\approx 10$) after the quench. 
For all theoretical studies presented, to simplify the calculations, we assume that the initial and final 
traps are axially symmetric (but about different axis).
Specifically,
our calculations use
$\omega_{y,i}=\omega_{z,i}$,
$\omega_{x,i}/\omega_{z,i}=0.2098$,
$\omega_{z,i}/\omega_{z,f}=494.3$,
$\omega_{x,f}=\omega_{y,f}$, and
$\omega_{x,f}/\omega_{z,f}=10$.

We record the spatial correlations along the $z$-axis 
 [Fig.~\ref{fig1}(aiii)] that develop during the wave packet dynamics using a single-atom and state resolved imaging scheme~\cite{exp2}. 
Furthermore, we control the interaction strength by adiabatically adjusting a magnetic offset field in the vicinity of a broad Feshbach resonance located at around $690$~G~\cite{feshbach}.
This allows us to reach three distinct regimes via the quench, 
which are set by the role of a bound state in the system 
[Fig.~\ref{fig1}(b)]: 
In the case of a small negative $a_s$ 
[in units of the harmonic oscillator length $a_{\text{ho},z}=\sqrt{\hbar/(\mu\omega_{z,f})}$], 
the system is in the weakly-attractive regime 
where the quench projects onto a large number of nearly free particle eigenstates. 
For $a_s/a_{\text{ho},z}=-0.0203$,
the occupation $|c_0|^2$ of the lowest eigenstate of $H_{\text{rel},f}$
immediately after the quench is about $2\%$ 
(Table~S1~\cite{SupMat}).
For small positive $a_s$ (e.g. $a_s/a_{\text{ho},z} = 0.0474$), in contrast, 
the particles are deeply bound into a single molecular state both before and after the quench. 
In this work, 
we are particularly interested in the paradigmatic ``unitary'' regime~\cite{fermion_rmp,bloch_rmp}, 
where the three-dimensional scattering length is the largest length scale in the system or even diverges.
In this regime ($a_s/a_{\text{ho},z}=-4.64$), 
$|c_0|^2$ is of order $0.3$ 
(Table~S1~\cite{SupMat}).

Since the quench does not couple the relative and center-of-mass motions,
the
center-of-mass wave packet for $t>0$ simply performs breathing oscillations
at the characteristic time scales $T_{x/y/z}/2= \pi / \omega_{x/y/z,f}$.
The relative motion, in contrast, is non-trivial.
Since the energy $\langle E_{\text{rel}} \rangle$ of the $t>0$ 
wave packet in the relative degrees of freedom is much larger than the energy scales
set by the trapping frequencies of $H_{\text{rel},f}$ 
(Table~S1~\cite{SupMat}),
the dynamics in the relative degrees of freedom
involves many eigenstates
of $H_{\text{rel},f}$.
To illustrate this,
the circles in Fig.~\ref{fig1}(b) schematically show the 
occupation probabilities 
$|c_j|^2$, 
which are 
obtained by expanding the relative portion 
of the $t<0$ wave packet 
in terms of the eigenstates of $H_{\text{rel},f}$~\cite{Calarco, Calarco_2},
for three different $s$-wave scattering lengths.

\begin{figure}
\vspace*{+0.2cm}
\includegraphics[scale=0.5, width=8cm]{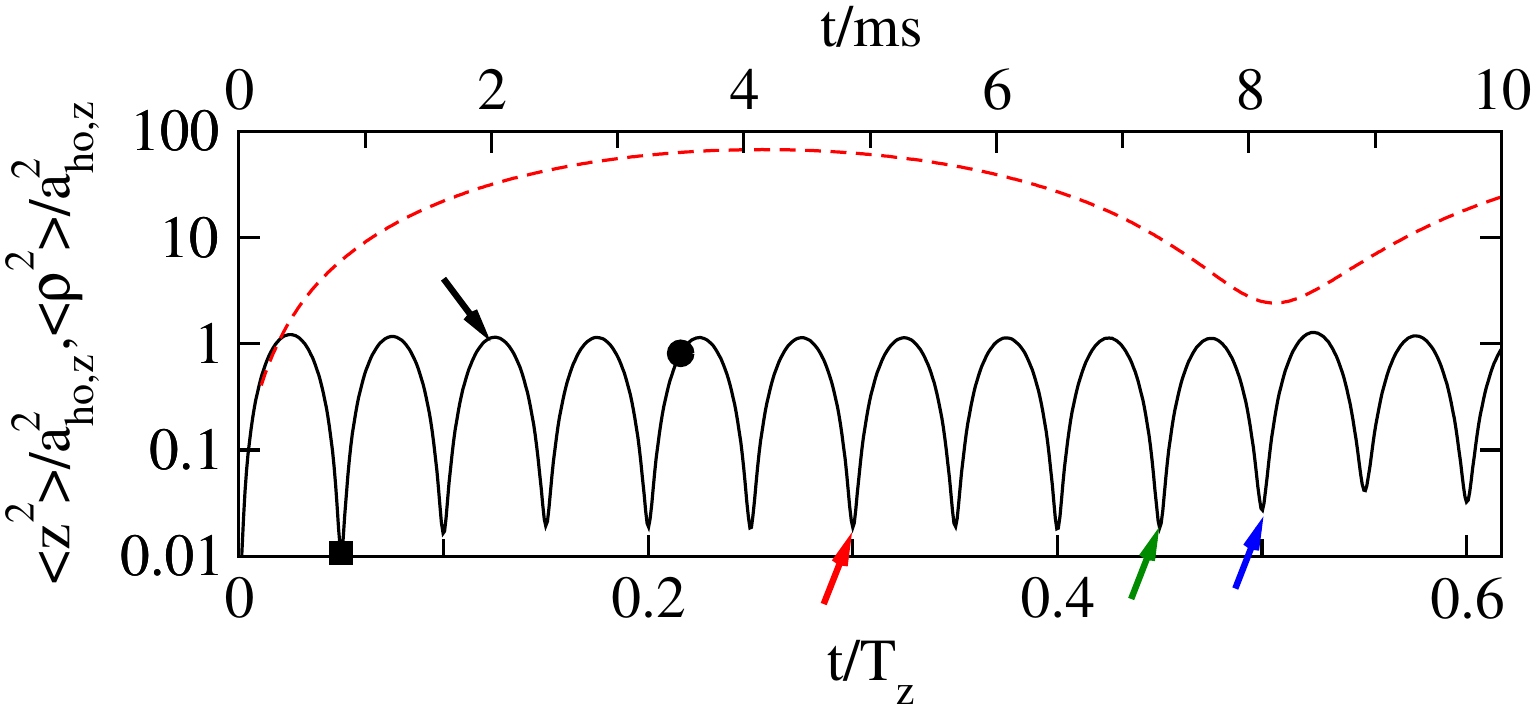}  
\vspace*{0.0cm}
\caption{(color online)
Expectation values of $\rho^2$ (solid line) and $z^2$ (dashed line) 
for $a_s=-4.64a_{\text{ho},z}$ 
as a function of time.
The circle, square, and arrows mark the times that are considered in
Figs.~\ref{fig2}, \ref{fig3}, and \ref{fig4}, respectively. 
}
\label{fig1p5}
\end{figure}

\begin{figure}
\vspace*{+0.0cm}
\includegraphics[scale=0.6]{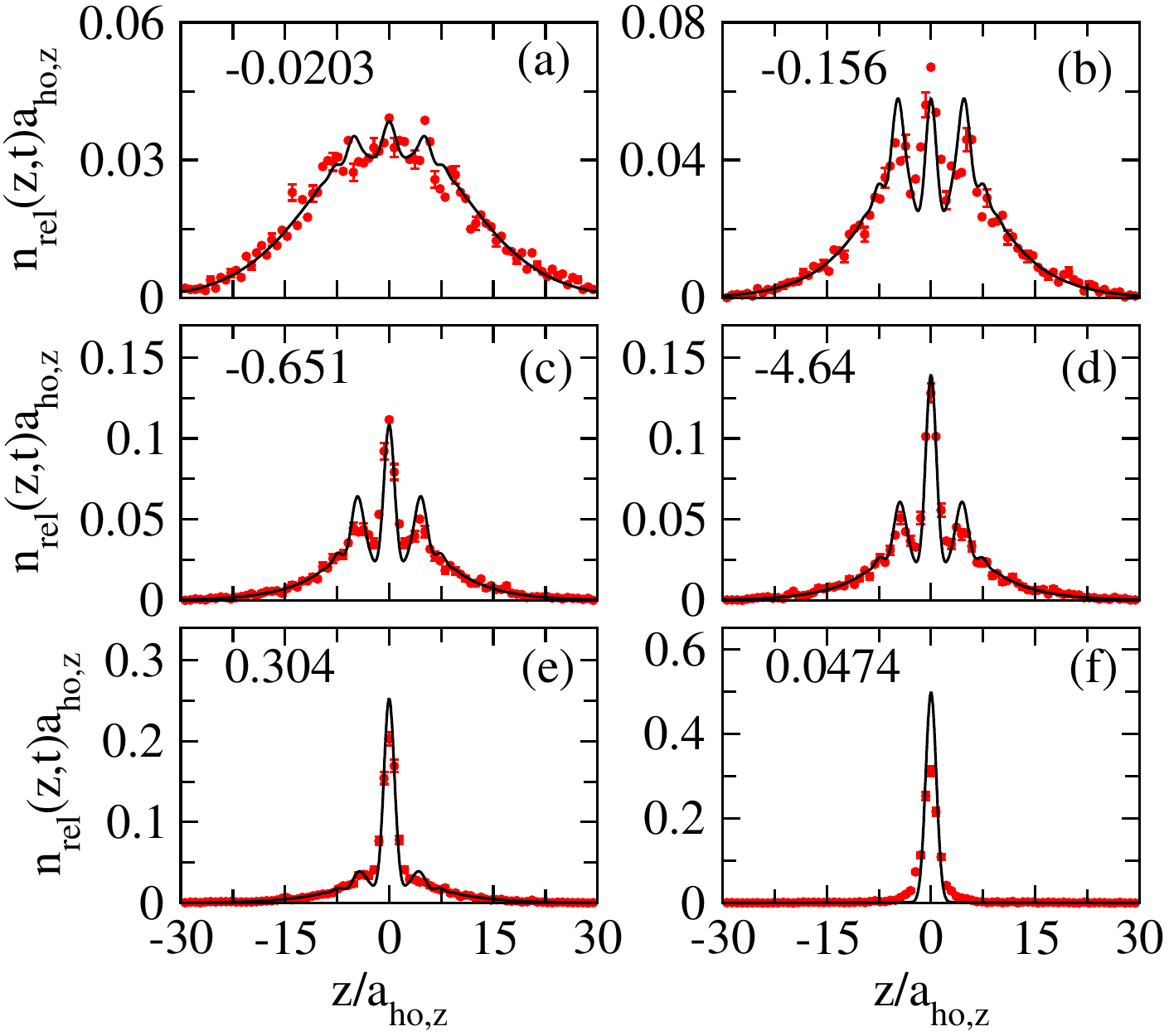}  
\vspace*{0.0cm}
\caption{(color online)
Experiment-theory comparison 
after four two-body collisions.
The relative densities $n_{\text{rel}}(z,t)$
are shown for $t=3.5~\text{ms}$ 
and six different scattering lengths.
The values of $a_s$ (in units of $a_{\text{ho},z}$) are reported in the upper left corner of each panel.  
Circles show experimental data and 
solid lines convolved theory results.
Typical error bars are shown for a subset of the experimental data.
Note the different $y$-scales in the panels.
}
\label{fig2}
\end{figure}

\begin{figure}
\vspace*{+.0cm}
\includegraphics[scale=0.6]{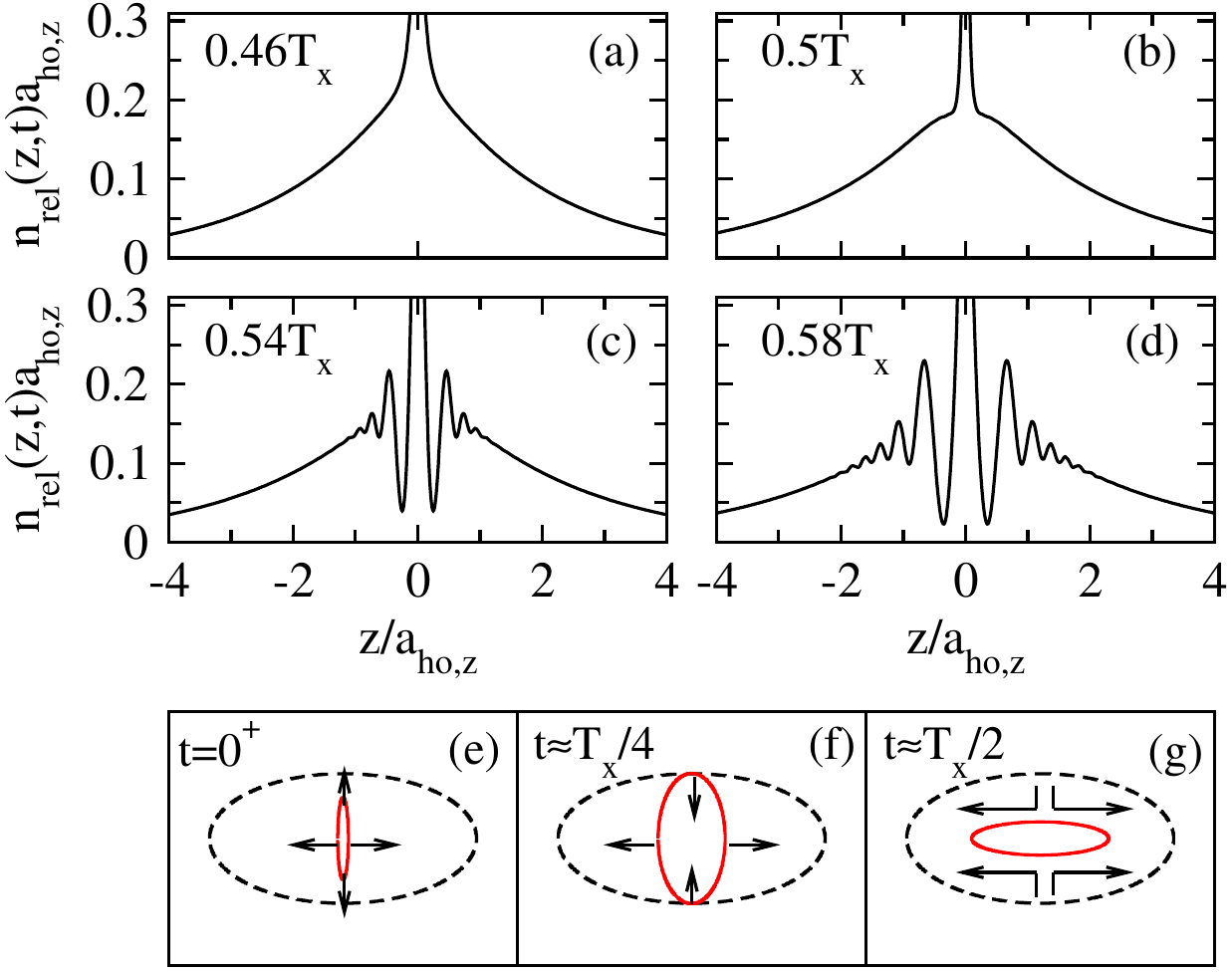}  
\vspace*{0.2cm}
\caption{(color online)
Development of fringe pattern in the relative density
during the first collision event
($t \approx T_x/2$) for $a_s = -0.651a_{\text{ho},z}$.
(a)-(d)
The lines show the theoretically determined 
unconvolved relative density $n_{\text{rel}}(z,t)$
for the times $t$ reported in the upper left corner of each panel.
(e)-(g) The red solid lines schematically show the wave packet at 
(e) $t=0^+$, (f) $t\approx T_x/4$, and (g) $t\approx T_x/2$.
The black dashed lines schematically show the equipotential lines of the 
final trap. 
The arrows schematically indicate the flux.  
}
\label{fig3}
\end{figure}

\begin{figure}
\vspace*{+.0cm}
\includegraphics[scale=0.53]{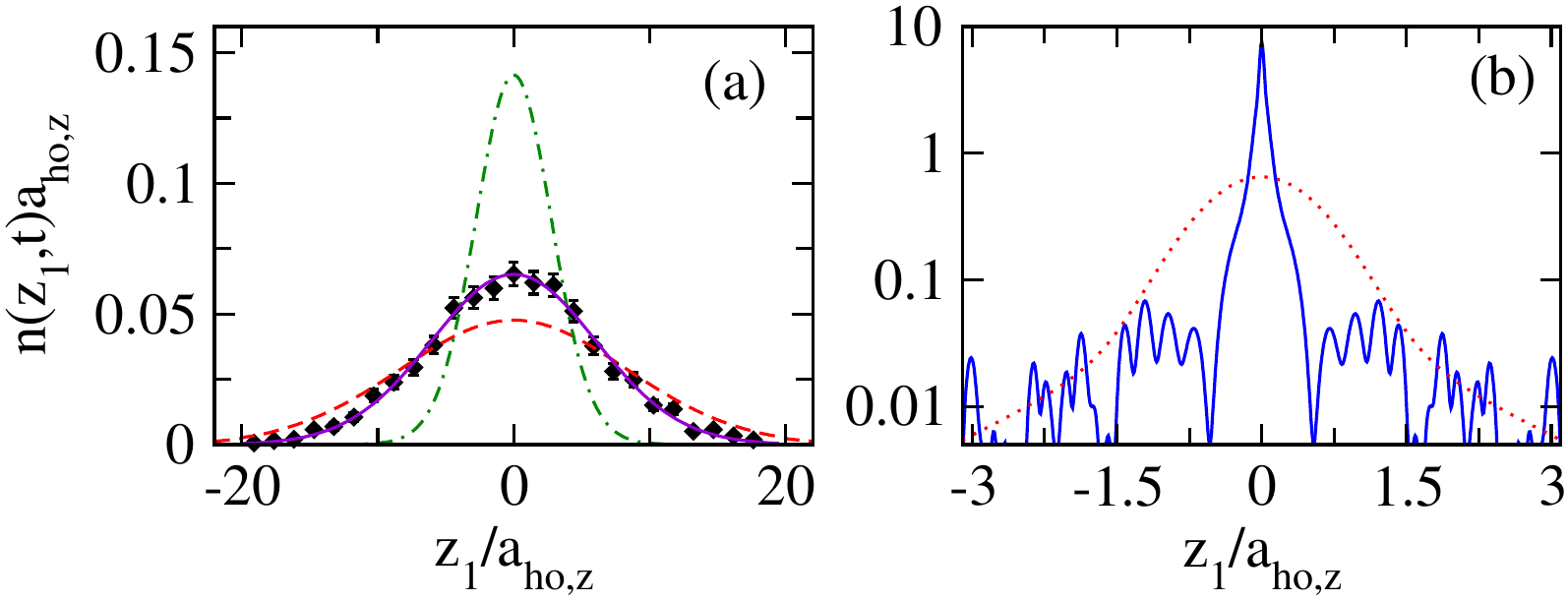}  
\vspace*{0.0cm}
\caption{(color online)
Single-particle densities for $a_s=-4.64a_{\text{ho},z}$.
(a) The black solid line and purple diamonds with error bars
show the convolved theoretical data and experimental results for $t=1.234T_x=2\text{ms}$;
the red dashed and green dot-dashed lines show the convolved theoretical data for $t=3T_x$ and $t=4.5T_x$, respectively.
(b) The blue solid and red dotted lines show, respectively, the unconvolved and convolved theoretical data for $t=5T_x=0.5T_z$.
}
\label{fig4}
\end{figure}

Figure~\ref{fig1p5} summarizes the dynamics for $a_s=-4.64a_{\text{ho},z}$
by displaying $\langle\rho^2\rangle$ and $\langle z^2\rangle$,
where 
$\rho^2=x^2+y^2$.
Both observables oscillate smoothly with time but at different frequencies. 
The times marked by a circle, a square, and arrows are discussed in more detail 
in Figs.~\ref{fig2}, \ref{fig3}, and \ref{fig4}, respectively.

Figure~\ref{fig2}
shows the relative density along the $z$-coordinate
for $t=3.5~\text{ms}=2.16T_x$, i.e., after four collisions 
(Fig.~\ref{fig1p5}), 
for six different 
$s$-wave scattering lengths.
The agreement between the experimental results
(circles) and the parameter-free theory results (solid lines)
is, except for Fig.~\ref{fig2}(f), very good.
The theoretical results shown in Fig.~\ref{fig2} are convolved with a Gaussian to account for the experimental resolution of  $4~\mu m = 0.542 a_{\text{ho},z}$. 
Interestingly, the relative densities shown in Figs.~\ref{fig2}(a)-\ref{fig2}(e)
contain
oscillatory structure or fringes,
which change notably
with the $s$-wave scattering length $a_s$,
on top
of a broad background.
The fringe pattern changes with time and we have found
no unique way to assign $t$-
and $z$-independent peak spacings for fixed
scattering length.
For the smallest positive
$s$-wave scattering length considered [Fig.~\ref{fig2}(f)], 
the initial
state is small compared to the
harmonic oscillator lengths of the final and initial traps
and the coefficient $|c_0|^2$ is large 
(Table~S1~\cite{SupMat}).
In this case, finite-range effects might need to
be accounted for to obtain quantitative agreement between theory and experiment.

The relative densities, shown in Figs.~\ref{fig2}(a)-\ref{fig2}(f),
reflect the evolution from a comparatively weakly-interacting regime,
in which the molecular state does not play a special role (small $|c_0|^2$),
to the strongly-interacting regime, 
where $|c_0|^2$ is appreciable but not dominant,
to the small molecular bound-state regime, 
where $|c_0|^2$ dominates. For large negative $a_s$ we observe a clear fringe pattern.
Since the wave packet in the relative coordinate 
would,
in the
absence of the scatterer, simply repeatedly expand and contract, 
the fringes have to be caused by scattering events.
Figure~\ref{fig3} 
shows the theoretically determined unconvolved relative density along $z$
for $a_s=-0.651a_{\text{ho},z}$ during the first scattering event,
i.e., for $t$ close to $t=T_x/2$.
At this time, 
$\langle\rho^2\rangle$ is quite small but $\langle z^2\rangle$
is comparatively large.
This implies that the majority of the wave packet is located away from the point scatterer. The snapshots in Fig.~\ref{fig3}
illustrate that the fringes emerge as a 
consequence of the scattering. 
A portion of the small-$\rho$
wave packet does not get reflected along the $\rho$-direction but instead
gets ``redirected'' to leave the small-$\rho$ region along the $z$-direction
[schematic in Fig.~\ref{fig3}(g)].
One can think of the scattering event as a cross-dimensional redistribution
of flux from the $\rho$- to the $z$-direction,
creating a newly emitted wave packet portion along the $z$-direction that
subsequently interferes with the ``background'' wave packet portion.
This process is repeated during subsequent scattering events 
($t\approx nT_x/2; n=2,3,4,...$),
leading to an increasingly complex fringe pattern
in the relative density along $z$ 
(see also Fig.~S3~\cite{SupMat}). 

\begin{figure}
\vspace*{+.2cm}
\hspace*{0cm}
\includegraphics[scale=0.6]{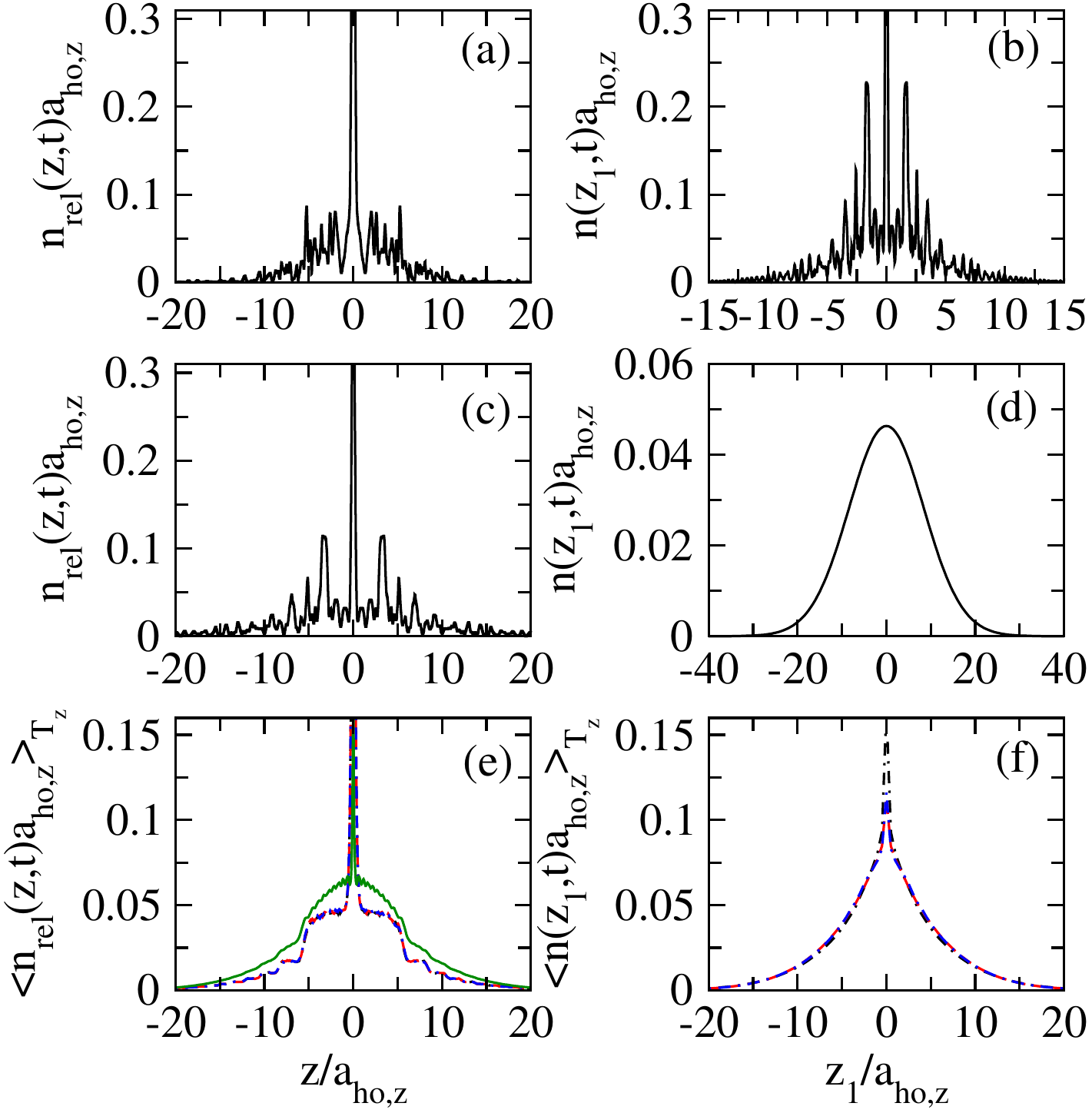}  
\vspace*{0.0cm}
\caption{(color online)
Unconvolved theory results for the long-time regime
for $a_s=-4.64a_{\text{ho},z}$.
The left and right columns show the relative 
and single-particle densities, respectively.
Panels~(a) and (b)
show snapshots for $t=4T_z$
while panels~(c) and (d) show  snapshots for $t=4.25T_z$.
Lines in panels~(e) and (f) show cycle-averaged observables
for
$n=1$, $10^2$, and $10^4$.
On the scale shown, 
the lines are essentially indistinguishable. 
For comparison,
the green solid line in panel (e)
shows the relative thermal density.
}
\label{fig5}
\end{figure}

Does the single-particle density,
an observable recorded frequently in cold atom experiments,
develop a fringe pattern?
The answer is yes but only for very short time periods over a length scale 
that is too small to be observed with the current experimental set-up.
Figure~\ref{fig4}(a) compares the experimental (diamonds) and convolved theoretical (solid line)
single-particle densities along $z$ 
for $t=2\text{ms}$ and $a_s=-4.64a_{\text{ho},z}$.
At this time, 
which corresponds to two oscillations of $\langle\rho^2\rangle$ 
(Fig.~\ref{fig1p5}),
the convolved single-particle density is smooth. 
It continues to be smooth for times $t<0.5T_z$
[red dashed and green dot-dashed lines in Fig.~\ref{fig4}(a)].
Since the size of the wave packet 
is much larger than the Gaussian convolution width $\sigma$, $\sigma=0.542a_{\text{ho},z}$,
the convolved and unconvolved single-particle density are 
indistinguishable on the scale shown in Fig.~\ref{fig4}(a).
The behavior of the single-particle density changes drastically  
when the wave packet is characterized 
by a small $\langle z^2\rangle$ and a small $\langle \rho^2 \rangle$.
For $t\approx 0.5T_z$,
the unconvolved single-particle density [solid line in Fig.~\ref{fig4}(b)] 
exhibits a fringe pattern.
The fringe pattern exists only for a short time period. 
For $t=0.501T_z$ (not shown), 
e.g., 
the oscillations are no longer visible.
Additionally, the limited spatial resolution smoothes out the fringe pattern of the single-particle density such that it cannot be observed in the experiment 
[dotted line in Fig.~\ref{fig4}(b)].
The fringe pattern in the single-particle density keeps 
``appearing'' and ``disappearing'' at larger times. 
Figure~\ref{fig5}(b) shows that
the single-particle density displays intricate fine structure
for $t=4T_z$ 
(corresponding to $\langle z^2\rangle\approx 0$). 
Figure~\ref{fig5}(d) shows that
no fringe pattern exists in the single-particle density
for $t=4.25T_z$
(corresponding to $\langle z^2\rangle\approx 70a_{\text{ho,z}}^2$).
Remarkably, the relative density along $z$ is characterized by notable fine structure for both times
[Figs.~\ref{fig5}(a) and \ref{fig5}(c)]. 

The discussion surrounding Figs.~\ref{fig2}-\ref{fig5}(d)
illustrates that collisions impact the single-particle and relative densities differently.
In particular, the relative density displays an increasingly large number of oscillations
with increasing time while the single-particle density is smooth for all times,
except for $t\approx n T_z/2$. 
Given the strong time dependence of the relative density,
we ask whether the system, in the large time limit, 
approaches thermal equilibrium.
The answer is, as is expected from Ref.~\cite{Olshanii_1},
that it does not. 
To gain insight into the long-time dynamics,
we analyze cycle-averaged observables, 
i.e.,
observables averaged over a period of length $T_z$
[from $t=nT_z$ to $t=(n+1)T_z$].
Lines in Figs.~\ref{fig5}(e) and \ref{fig5}(f) 
show the unconvolved cycle-averaged relative and single-particle densities 
$\langle n_{\text{rel}}(z,t)\rangle_{T_z}$ and $\langle n(z_1,t)\rangle_{T_z}$,
respectively, for $a_s=-4.64a_{\text{ho},z}$ and $n=1$, $10^2$, and $10^4$.
In both panels, the three curves are indistinguishable on the scale shown.
Thus, despite the intricate dynamics within each cycle,
the cycle-averaged observables display essentially no dynamics. 
The reason for this is that the normalized nearest neighbor energy spacings 
are rather sharply peaked around 1 
(Fig.~S1~\cite{SupMat}). 
We emphasize that this behavior is also observed for other scattering lengths.
The close to frozen cycle-averaged relative density for large $n$ indicates a lack of thermalization. 
Indeed, the thermal relative density [green solid line in Fig.~\ref{fig5}(e)]
differs visibly from the calculated cycle-averaged relative densities. 

In summary, 
we have presented a joint theoretical-experimental study that investigated the wave packet
dynamics of two ultracold atoms following a ``violent'' trap quench,
which leads to the occupation of many eigenstates of the post-quench Hamiltonian.
Following the quench, two-body collisions,
through their effect on the structural observables, were observed. 
The excellent agreement between the experimental and theoretical data together 
with the time-resolved single-atom detection with high spatial resolution 
makes the system a promising candidate for future dynamical studies,
which are aimed at addressing questions related to thermalization,
state engineering, chaos, and integrability. 
The set-up also promises to be a fertile playground for testing hydro-dynamical formulations~\cite{hydro_1, hydro_2, hydro_3}, 
which can potentially be used to simulate the dynamics of few- and many-body systems.
Quantitative tests of the hydrodynamics theory with ultracold atoms may yield insights into why the dynamics of quark gluon plasmas seems, 
somewhat surprisingly, to be governed by hydrodynamic equations~\cite{gluon_1, gluon_2}. 

\section{Acknowledgement}
QG and DB gratefully acknowledge support by the National Science Foundation through
grant number PHY-1806259. This
work has partially been supported by the ERC consolidator grant
725636, the Heidelberg Center for Quantum Dynamics,
and the DFG Collaborative Research Centre SFB 1225
(ISOQUANT). PMP acknowledges funding from European Union’s Horizon 2020 programme under the Marie Sklodowska-Curie grant agreement No. 706487 and the Daimler and Benz Foundation. AB acknowledges funding from the International Max-Planck Research School (IMPRS-QD). 
This work used the OU
Supercomputing Center for Education and Research
(OSCER) at the University of Oklahoma (OU).

\pagebreak
\begin{center}
\textbf{\large Supplemental material: Density Oscillations Induced by Individual Ultracold Two-Body Collisions}
\end{center}
\setcounter{equation}{0}
\setcounter{figure}{0}
\setcounter{table}{0}
\setcounter{page}{1}
\makeatletter
\renewcommand{\theequation}{S\arabic{equation}}
\renewcommand{\thefigure}{S\arabic{figure}}
\renewcommand{\bibnumfmt}[1]{[S#1]}
\renewcommand{\citenumfont}[1]{S#1}

\author{Q. Guan}
\affiliation{Homer L. Dodge 
Department of Physics and Astronomy, The University of Oklahoma, 
440 West Brooks Street,
Norman, Oklahoma
73019, USA}

\author{V. Klinkhamer}
\author{R. Klemt}
\author{J. H. Becher}
\author{A. Bergschneider}
\altaffiliation[Current address: ]{Institute of Quantum Electronics, ETH Zurich,
CH-8093 Zurich, Switzerland}
\author{P. M. Preiss}
\author{S. Jochim}
\affiliation{Physics Institute, Heidelberg University,
Im Neuenheimer Feld 226, 69120 Heidelberg, Germany}

\author{D. Blume}
\affiliation{Homer L. Dodge 
Department of Physics and Astronomy, The University of Oklahoma, 
440 West Brooks Street,
Norman, Oklahoma
73019, USA}

\date{\today}

\maketitle

\renewcommand{\theequation}{S\arabic{equation}}
\renewcommand{\thefigure}{S\arabic{figure}}
\renewcommand{\thetable}{S\arabic{table}}
\renewcommand{\bibnumfmt}[1]{[S#1]}
\renewcommand{\citenumfont}[1]{S#1}

\section{Experimental details}
The experiments described in this paper are all performed with two $^6$Li atoms in two of the three lowest hyperfine states, denoted by $\ket{1}$ and $\ket{3}$. 
The experimental procedure starts from a balanced, degenerate mixture of these two hyperfine states, which we obtain after evaporating the gas close to the Feshbach resonance at around
$690~\text{G}$. 
After evaporation we increase the magnetic field to above the Feshbach resonance to prepare a weakly-attractive Fermi gas [left side of Fig.~\ref{fig_S1}(a)].
We then transfer the atoms into an optical tweezer and spill to the motional ground state~\cite{exp1_S}. This state is adiabatically connected to the molecular state for positive $a_{\text{ho},z}/a_s$. 
We reach a ground state fidelity of approximately $97\%$. 
To study the two-atom system at different interaction strengths, we adiabatically ramp the magnetic field in 
$\sim 100\text{ms}$ 
to values between 
$665~\text{G}$ 
and 
$900~\text{G}$ 
(see Table~\ref{tab1}).
Afterwards we slowly ramp on the weak crossed-beam optical dipole trap. 
As the dipole trap provides a much weaker confinement compared to the tight tweezer it does not significantly affect the initial state. 
At $t=0$ we instantaneously switch off the tweezer trap and allow the wave packet to expand in the crossed-beam dipole trap. 
We keep the magnetic field constant during expansion. 
After expansion in the weak trap we apply a free-space single-atom imaging scheme~\cite{exp2_S} to obtain a spin-resolved image of the two atoms after various times of flight, 
which is spatially resolved along the $z$-axis, while integrating out the other two axes.

\section{Zero-range framework}

\begin{figure}
\vspace*{+.0cm}
\includegraphics[scale=0.5]{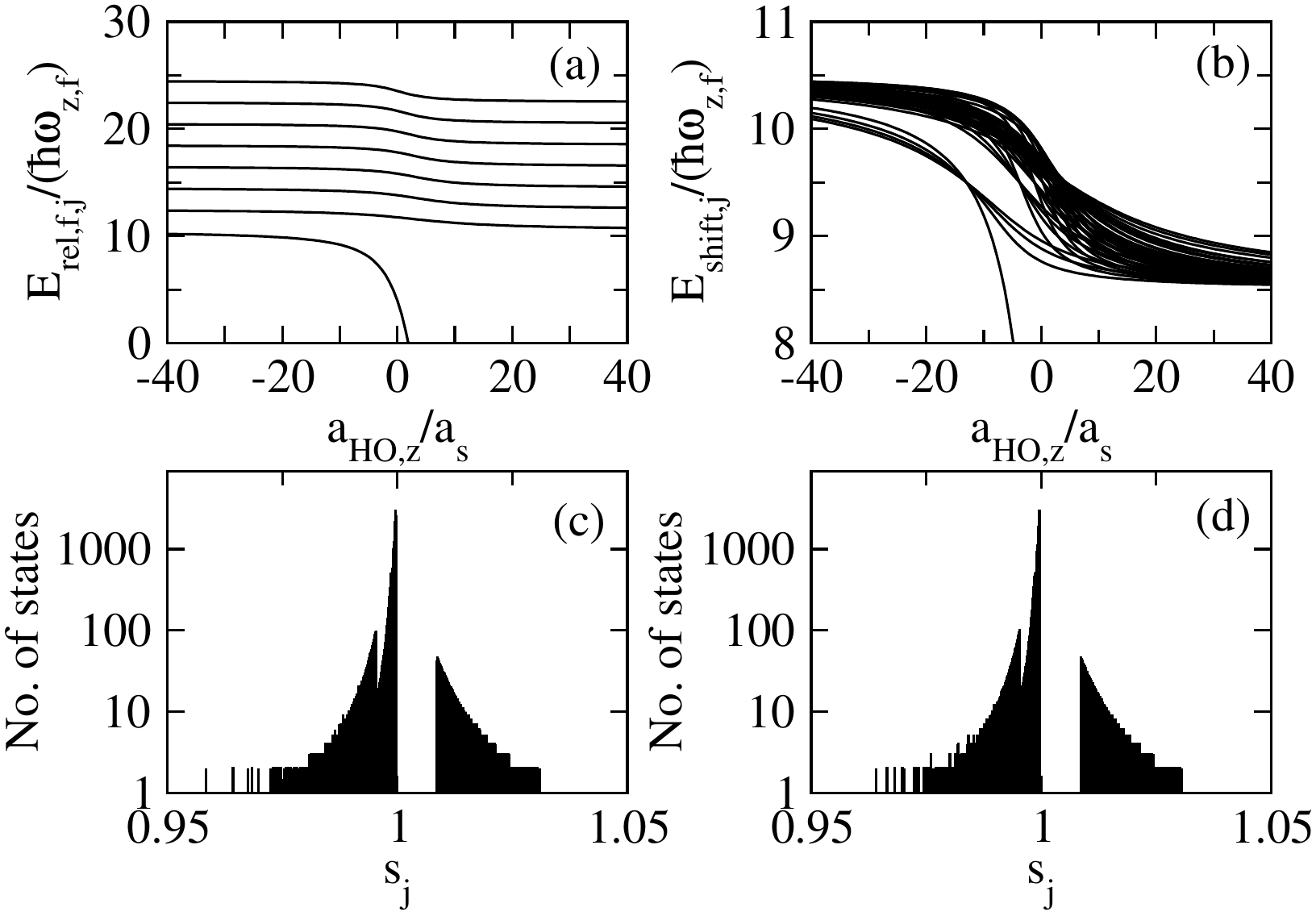}  
\vspace*{0.2cm}
\caption{
Characteristics of relative $m_l=0$ energy spectrum for $\eta=10$.
(a) The solid lines show the eight lowest relative $m_l=0$ eigenenergies $E_{\text{rel},f,j}$
as a function of $a_{\text{ho},z}/a_s$.
(b) The solid lines show the shifted energies $E_{\text{shift},j}$
for the $40$ lowest relative $m_l=0$ eigenenergies.
The histograms in (c) and (d) show the number of states with normalized nearest neighbor spacing
$s_j$ for $a_s=\infty$ and $a_s=0.304a_{\text{ho},z}$, respectively.
Note the logarithmic scales of the $y$-axis.
}
\label{fig_S1}
\end{figure}

The eigenenergies and eigenfunctions of $H_{\text{rel},i}$ 
and $H_{\text{rel},f}$ 
are obtained using the zero-range framework
developed in Ref.~\cite{Calarco_S}.
For integer aspect ratio $\eta$,
as considered in our calculations,
the determination of the eigenenergies involves the evaluation of a finite number of hypergeometric functions.
The calculations can be performed efficiently by employing an iterative scheme.
As an example, Fig.~\ref{fig_S1}(a)
shows the resulting low-lying portion of the relative energy spectrum 
with $m_l=0$
as a function of $a_{\text{ho},z}/a_s$
for $\eta=10$.
The $m_l$ quantum number, which is associated with the relative orbital angular momentum
operator $L_z$ along the $z$-axis,
is a good quantum number for the $t>0$ Hamiltonian $H_{\text{rel},f}$.
Thus,
the different $m_l$ channels that the initial state is projected onto
evolve independently.
To visualize the energy level distribution of the entire $m_l=0$ spectrum,
we shift the $j$-th energy level down by $2j\hbar\omega_{z,f}$.
The resulting energies $E_{\text{shift},j}$,
$E_{\text{shift},j}=E_{\text{rel},f,j}-2j\hbar\omega_{z,f}$,
are shown in Fig.~\ref{fig_S1}(b) as a function of $a_{\text{ho},z}/a_s$.
It can be seen that all the energies are 
folded into a single ``energy band''.
On the negative scattering length side,
the energy band is split into two sub-bands 
that are separated by a ``gap''.
This indicates that the system supports two different types of excitations:
excitations that lie predominantly along
the $z$-direction and excitations that lie predominantly 
along the $\rho$-direction. 
For all cases considered in this work, 
the $m_l=0$ channel contributes more than $95\%$ to the total weight.
The results presented account for the $m_l=0$ and finite $m_l$ components.
 
Figures~\ref{fig_S1}(c)-\ref{fig_S1}(d)
show the 
distribution of the nearest neighbor spacings $s_j$, $s_j=(E_{j+1}-E_j)/\bar{s}$, 
where $\bar{s}$ denotes the average of the nearest neighbor spacings,
for $a_s=\infty$ and $a_s=0.304a_{\text{ho},z}$, respectively, for $\eta=10$.
The lowest 100,021 energy levels with $m_l=0$ are included in Figs.~\ref{fig_S1}(c) and \ref{fig_S1}(d).
The nearest neighbor distributions peak at $s_j=1$ and fall off rapidly for smaller and larger $s_j$.
Importantly, even the 
energy spectrum
for $\eta=10$ and
$a_s=\infty$ contains more than one distinct energy spacing
despite the fact that the point scatterer does not set a length scale in this case.
The reason is that the system is characterized by two length scales, namely
the harmonic oscillator lengths in the axial and transverse directions.
The situation here is thus different from 
the strictly one-dimensional system and the spherically
symmetric three-dimensional
system with infinite coupling constants, which are characterized by a 
single harmonic oscillator length 
and equidistant energy level spacings~\cite{Busch_S}.
The distributions shown in Figs.~\ref{fig_S1}(c)-\ref{fig_S1}(d)
are notably different from
a GUE (Gaussian unitary ensemble) distribution, 
indicating non-chaotic behavior~\cite{GUE_1_S, GUE_2_S}.
The GUE of random matrices has been used extensively in the literature to analyze quantum chaotic behavior. 
It is also worthwhile pointing out
that the distributions are not Poissonian and  
different from those for a Seba billiard~\cite{Seba_S}.

To time evolve the initial state under the post-quench Hamiltonian $H_f$,
the relative portion of the initial state is expanded 
in terms of the relative eigenstates of $H_{\text{rel},f}$~\cite{Calarco_S}.
Provided the expansion coefficients are known,
the time dynamics amounts to keeping track of the phase factors.
Time-dependent structural observables are obtained by
combining the relative and center-of-mass portions of the
wave packet and integrating numerically over a subset of the spatial degrees of freedom.
To obtain accurate results for small and large
interparticle spacings, the relative wave packet is constructed using non-linear grids in
$\rho$ and $z$.

To obtain the expansion coefficients $c_j$
right after the quench, we
numerically calculate the overlap integrals between the initial state $\psi_{\text{rel},i}$ and 
the $j$-th eigenstate $\psi_{\text{rel},f,j}$ of $H_{\text{rel},f}$.
Since $\psi_{\text{rel},i}$ has the most significant overlap with $\psi_{\text{rel},f,j}$ at small distances,
special care has to be taken to obtain an accurate representation of the
$\psi_{\text{rel},f,j}$ at small $\rho$ and $z$.
As pointed out in Ref.~\cite{Calarco_S}, the convergence of the
infinite sums in Eq.~(56) of Ref.~\cite{Calarco_S} is quite slow
for small $\rho$ and/or $z$.
To deal with this, we adjust the number of terms included in the expansion
depending on the $\rho$ and $z$ values considered.
Table~\ref{tab1} lists the occupation probabilities $|c_0|^2$ for the magnetic field
strengths considered in this work.

\begin{table}
\caption{
The first two columns show the magnetic fields $B$ and the corresponding
$s$-wave scattering lengths considered in this work.
Column 3 reports the occupation of the lowest eigenstate of 
$H_{\text{rel},f}$ after the quench.
Column 4 reports the relative energy after the quench.
}
\begin{ruledtabular}
\begin{tabular}{llll} 
$B/\text{G}$ & 
$a_s/a_{\text{ho},z}$ & 
$|c_0|^2$ &
$\langle E_{\text{rel}} \rangle / (\hbar \omega_{z,f})$ \\ 
\hline
900 &  $-0.0203$ & $0.0207$ & $241.2$    \\
700 &  $-0.156$  & $0.0843$ & $143.9$    \\
692 &  $-0.651$  & $0.212$  & $111.4$    \\
690 &  $-4.64$   & $0.278$  & $101.0$    \\
685 &  $0.304$   & $0.516$  & $69.82$    \\
665 &  $0.0474$  & $0.979$  & $-209.9$   \\
\end{tabular}
\end{ruledtabular}
\label{tab1}
\end{table}

\section{Convolved and unconvolved distributions}
As discussed in the main text, the experimental resolution
when measuring the position
coordinate $z_j$ of the $j$-th atom along the axial direction
is $\sigma=4$~$\mu m$.
The convolved theoretically determined single-particle and relative
densities are based
on
\begin{align}
\label{S1}
n_{\text{con}}(z_1,t)= \nonumber \\
\int_{-\infty}^{\infty}n_{\text{uncon}}(z_1',t)\exp\left[-\frac{(z_1'-z_1)^2}{2\sigma^2}\right]dz_1',
\end{align}
and
\begin{align}
\label{S2}
n_{\text{rel},\text{con}}(z,t)= \nonumber \\
\int_{-\infty}^{\infty}n_{\text{rel},\text{uncon}}(z',t)\exp\left[-\frac{(z'-z)^2}{2(\sqrt{2}\sigma)^2}\right]dz'
\end{align}
where the subscripts ``con'' and ``uncon'' refer to ``convolved'' and ``unconvolved'', respectively.
In the main text,
the subscripts ``con'' and ``uncon'' are dropped for notational simplicity.
The widths of the Gaussian in Eqs.~(\ref{S1}) and (\ref{S2}) differ since the single-particle resolution $\sigma$
implies a resolution of $\sqrt{2}\sigma$ for the relative coordinate $z$.

\begin{figure}
\vspace*{+.0cm}
\includegraphics[scale=0.6]{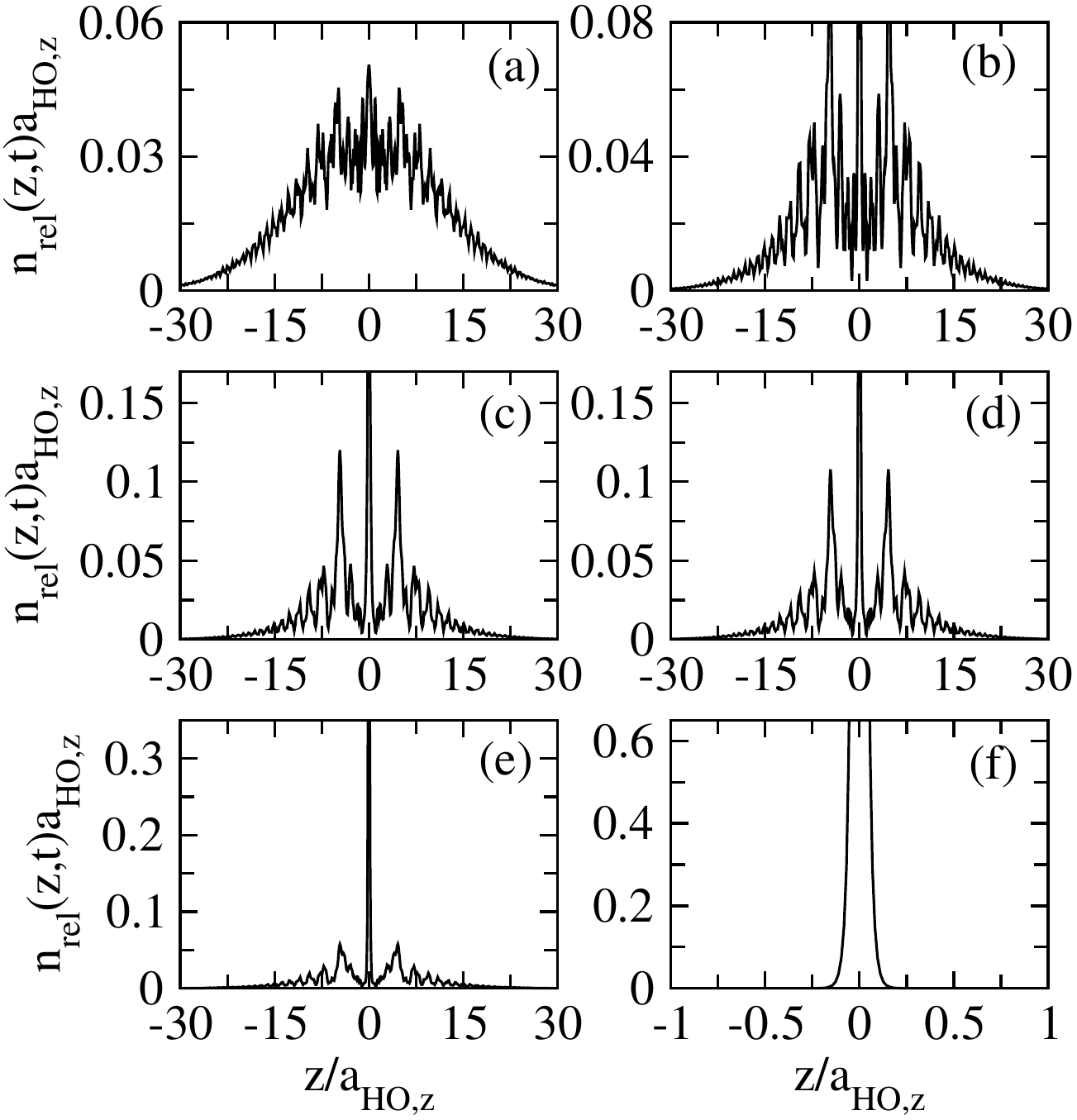}  
\vspace*{0.2cm}
\caption{
Unconvolved theory results 
corresponding to the convolved theory results shown in Fig.~3 
of the main text. 
Note that the extend of the axis is, except for the $x$-axis of panel (f),
the same as in Fig.~3 of the main text.
}
\label{fig_S3}
\end{figure}

Figure~\ref{fig_S3} shows the unconvolved data corresponding to
Fig.~3 of the main text.

\section{Time evolution of relative density}

\begin{figure}
\vspace*{+.0cm}
\includegraphics[scale=0.3, width=7.5cm]{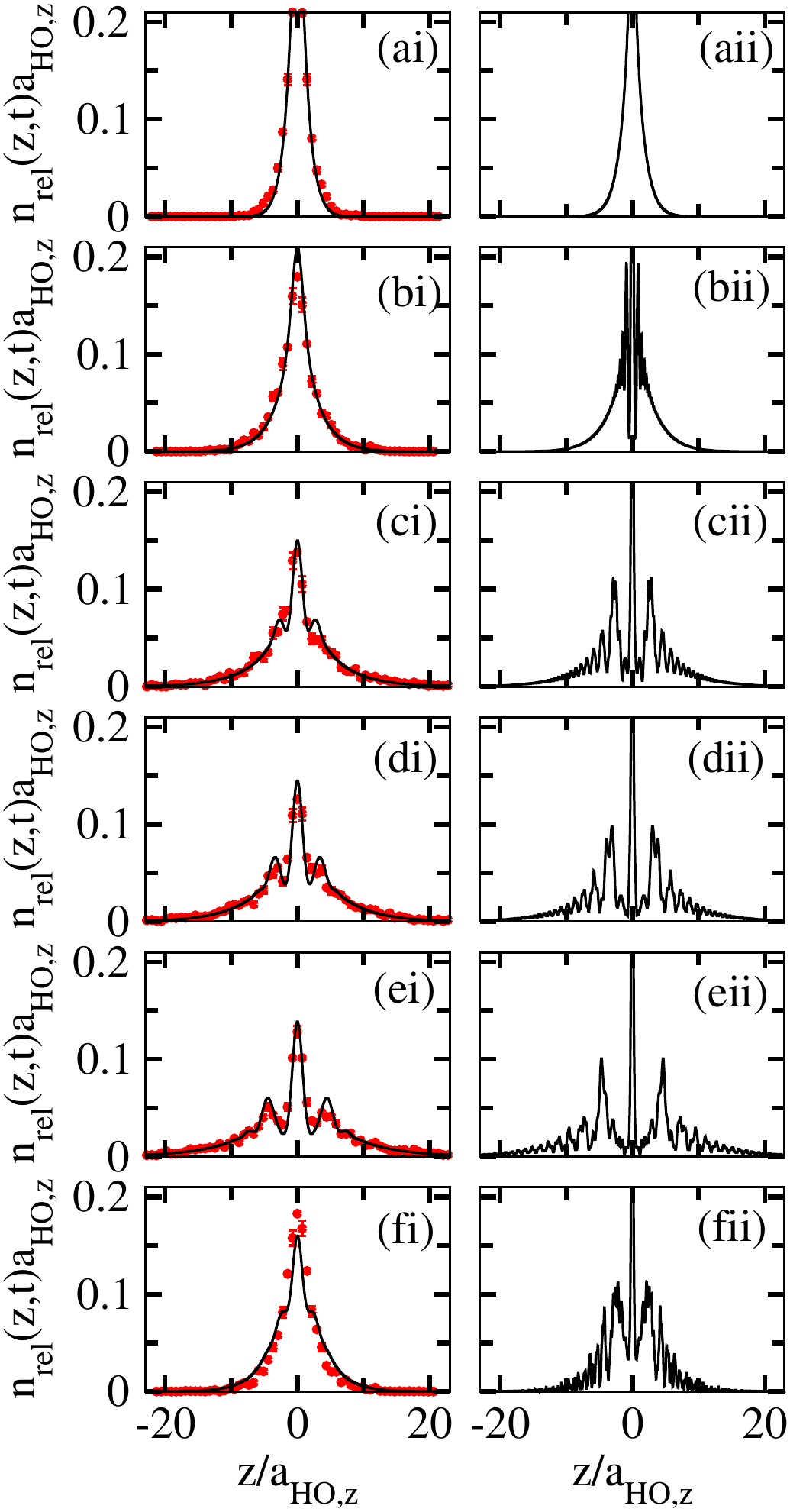}  
\vspace*{0.2cm}
\caption{(color online)
Convolved and unconvolved relative densities at various times for $a_s=-4.64a_{\text{ho},z}$.
Panels (a)-(f) correspond to $t=0.5\text{ms}$, $t=1\text{ms}$,
$t=2\text{ms}$, $t=2.5\text{ms}$, $t=3.5\text{ms}$, and $t=7\text{ms}$, respectively.
The system has undergone
0, 1, 2, 3, 4, and 8 collisions for panels~(a)-(f).
The left column compares the experimental results (red circles)
and convolved theoretical results (solid line).
Typical error bars are shown for a subset of the experimental data.
The right column shows the unconvolved theoretical results.
}
\label{fig_S2}
\end{figure}

Figure~\ref{fig_S2} shows the relative densities for $a_s=-4.64a_{\text{ho},z}$ and
six different times.
The experimental data and convolved theoretical results are in excellent agreement
(see the left column of Fig.~\ref{fig_S2}).
The right column of Fig.~\ref{fig_S2} shows the unconvolved theoretical results. 
For $t=0.5\text{ms}$ [Figs.~\ref{fig_S2}(ai) and \ref{fig_S2}(aii)],
no fringe pattern exists in either the convolved or unconvolved data
because collisions have not yet happened.
For $t=1\text{ms}$ [Figs.~\ref{fig_S2}(bi) and \ref{fig_S2}(bii)],
the system has undergone one collision. 
While the unconvolved data does show 
a fringe pattern at this time [see Fig.~\ref{fig_S2}(bii)],
the convolved data does not [see Fig.~\ref{fig_S2}(bi)]. 
For $t=2\text{ms}$ [Figs.~\ref{fig_S2}(ci) and \ref{fig_S2}(cii)],
the system has undergone two collisions. 
For this time, two types of oscillations can be seen on top of each other in the unconvolved data;
one has already propagated out and the other exists only at small $z$ [see Fig.~\ref{fig_S2}(cii)].
After two collisions, the fringe pattern is visible in the convolved data [see Fig.~\ref{fig_S2}(ci)].   
For $t=2.5\text{ms}$ [Figs.~\ref{fig_S2}(di) and \ref{fig_S2}(dii)] 
and $t=3.5\text{ms}$ [Figs.~\ref{fig_S2}(ei) and \ref{fig_S2}(eii)],
the system has undergone three and four collisions, respectively.
For these times, the unconvolved relative densities show more and more fine structure
[see Figs.~\ref{fig_S2}(dii) and \ref{fig_S2}(eii)].
Correspondingly, 
the contrast of the fringe pattern in the convolved relative density increases 
and small wiggles can be seen in addition to the three main peaks 
(one central peak at $z=0$ and two side peaks at $z\approx \pm 4a_{\text{ho},z}$).  
For $t=7\text{ms}$ [Figs.~\ref{fig_S2}(fi) and \ref{fig_S2}(fii)],
the system has undergone eight collisions. 
It can be seen that
the unconvolved relative density is governed by multiple frequencies.
Since the oscillations are occuring on a small length scale, 
the finite spatial resolution is not sufficient to resolve
the fringe pattern in the convolved data, i.e.,
Fig.~\ref{fig_S2}(fi) shows only a hint of a shoulder.

\section{Finite-range interactions}
\begin{figure}
\vspace*{+.0cm}
\includegraphics[scale=0.6]{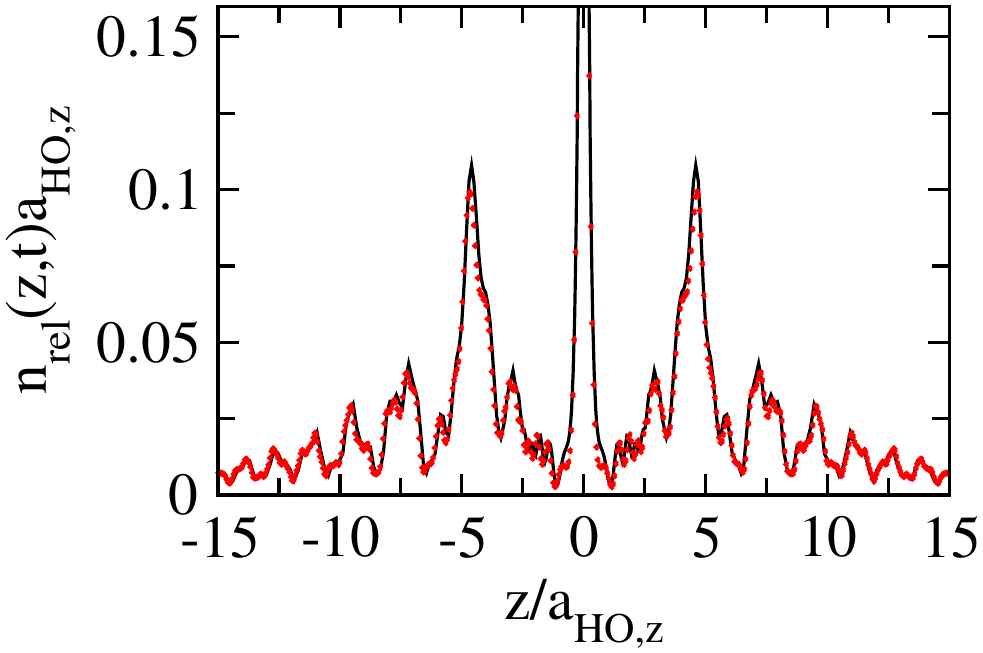}  
\vspace*{0.2cm}
\caption{(color online)
Zero-range versus finite-range results for $t=3.5\text{ms}$ and $a_s=-4.64a_{\text{ho},z}$.
The black solid line shows the unconvolved relative density along the $z$-coordinate, for the zero-range potential for $\eta=10$.
For comparison, the red circles show the unconvolved relative density along the $z$-coordinate for the Gaussian potential with $r_0=0.00495a_{\text{ho},z}$
for $\eta=10.01$.
The initial state is prepared identically in both cases,
using the trap frequencies $\omega_{y,i}=\omega_{z,i}=30.5\text{kHz}$, and $\omega_{x,i}=6.4\text{kHz}$ 
(see the main text).
}
\label{fig_S4}
\end{figure}

To analyze the role of effective range corrections,
we compare results for the zero-range potential
and an attractive 
Gaussian potential with range $r_0=0.00495a_{\text{ho},z}$, 
for which the depth is
adjusted to dial in the desired $s$-wave scattering length.
We restrict ourselves to finite-range potentials that support zero or one free-space $s$-wave bound state.
For the Gaussian potential, the time evolution is performed by 
separately propagating each
$m_l$ component of the relative portion of the wave packet.
We find that the lowest few $m_l$ provide a good description 
of the initial state. 
Each $m_l$ component is, in turn, expanded in terms of 
the product of
$m_l$-dependent radial functions $R_{lm_l}(r)$ and
spherical harmonics $Y_{lm_l}$,
where the relative orbital angular momentum
quantum number $l$ runs over even values ($l=0,2,4,...$).
The resulting set of coupled radial equations is 
propagated in time using the techniques
described in Refs.~\cite{chebyshv_S}.
Since the dynamics leads to the occupation of many partial waves, the
convergence with respect to $l$ needs to be checked carefully.
The results 
shown in Fig.~\ref{fig_S4} include $|m_l|$ values up to 4
and $l$ values up to $500$.

Figure~\ref{fig_S4} compares the relative density along $z$
for $t=3.5\text{ms}$ and $a_s=-4.64a_{\text{ho},z}$.
The aspect ratio $\eta$ is set to 10 for 
the zero-range interaction (black solid line) and to 10.01 for the Gaussian interaction (red circles). 
The agreement between the line and the symbols is excellent. 
This shows that the relative density along $z$ is insensitive to the short-range details of the interactions and
that a small change in the aspect ratio has a negligible effect on the
short-time dynamics.
We expect the latter to be true quite generally. 
The former, in contrast, is
only expected to hold if $|a_s|$ is much larger than $r_0$.

\section{Relative thermal density}
To obtain the relative thermal density shown in Fig.~6(e) of the main text, 
we define an effective relative temperature $T_{\text{rel}}^{\text{eff}}$ through
\begin{eqnarray}
\label{rel_T}
\langle E_{\text{rel}}\rangle = \sum_j|b_j|^2E_{\text{rel},f,j},
\end{eqnarray}
where $j$ runs over all eigenstates of $H_{\text{rel},f}$ and the $|b_j|^2$ are the Boltzmann factors,
\begin{eqnarray}
\label{boltzmann}
|b_j|^2=Z^{-1}\exp\left(-\frac{E_{\text{rel},f,j}}{k_b T_{\text{rel}}^{\text{eff}}}\right),
\end{eqnarray}
with $Z$ chosen such that the normalization 
\begin{eqnarray}
\label{nomalization}
\sum_j|b_j|^2=1
\end{eqnarray}
is fulfilled.
Here, $\langle E_{\text{rel}}\rangle$
is the relative energy after the quench (see Table~\ref{tab1}),
and $k_b$ the Boltzmann constant.
Physically, the effective temperature is defined assuming that the system in the relative degrees of freedom is described by a canonical ensemble. 
The relative thermal density is then given by 
\begin{eqnarray}
\label{thermal}
n_{\text{thermal}}(z)=2\pi \sum_j \int_0^{\infty}|b_j|^2|\psi_{\text{rel},f,j}(\vec{r})|^2\rho d\rho.
\end{eqnarray}

\section{Comparison with dynamics of strictly one-dimensional system}
An important question is the following: Does a strictly one-dimensional system
exhibit analogous dynamics or does the three-dimensional character of
our set-up play a crucial role?
We find that the observed short-time dynamics
depends notably on the three-dimensional character of
the set-up
despite the fact that the trap is highly-elongated.
To arrive at this conclusion,
we performed calculations for a strictly one-dimensional system~\cite{Busch_S}.
Specifically, we 
consider a quench of a
strictly one-dimensional two-particle system
with contact interaction of strength $g_{\text{1D}}$.
We prepare the system in its ground state
at $t<0$ and consider the situation
where the trap frequency is reduced by a factor of 100
at time $t=0$
(we have checked that a reduction of the trap frequency by a factor of 10
yields qualitatively similar results).

\begin{figure}
\vspace*{+0cm}
\includegraphics[scale=0.58]{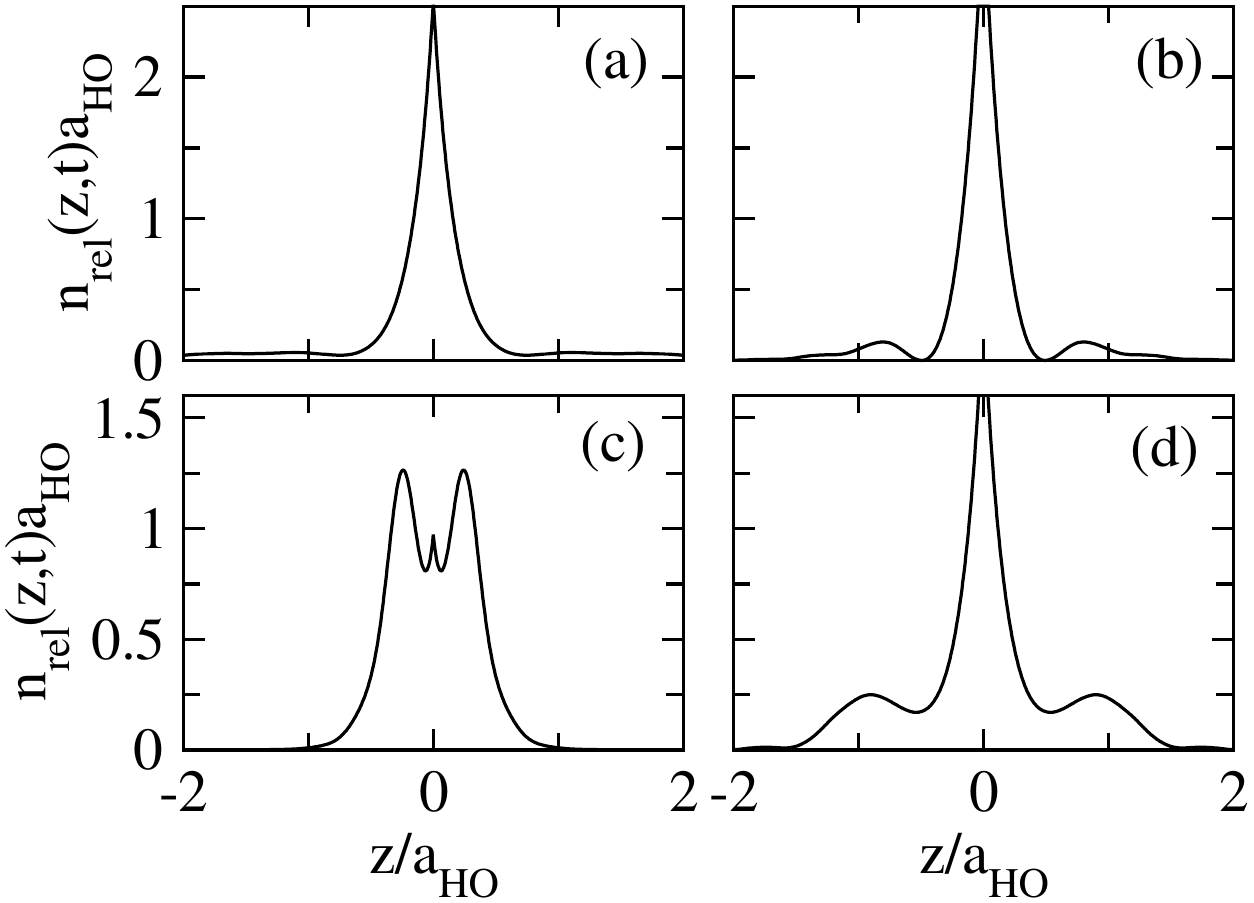}  
\vspace*{0.2cm}
\caption{
Dynamics for strictly one-dimensional system with zero-range interactions. 
The solid lines show 
the relative densities for $g_{\text{1D}}/(a_{\text{ho}}\hbar\omega)=-4.19$
and
(a) $t=0.48T$,
(b) $t=0.49T$,
(c) $t=0.5T$,
and
(d) $t=0.51T$,
respectively.
}
\label{fig_S5}
\end{figure}

Figure~\ref{fig_S5} shows the unconvolved relative density
for $g_{\text{1D}}/(a_{\text{ho}}\hbar\omega)=-4.19$ and 
$t=0.48T$, $t=0.49T$, $t=0.5T$, and $t=0.51T$,
i.e., just before and after the system has undergone the first collision.
Here, $\omega$ denotes the trap frequency and
$a_{\text{ho}}$ is the corresponding trap length scale, 
$a_{\text{ho}}=\sqrt{\hbar/(\mu\omega)}$.
$T$ is the oscillation period, $T=2\pi/\omega$. 
Comparison of 
Fig.~\ref{fig_S5} and Fig.~3 of the main text
shows 
that fewer fringes are formed per scattering for the strictly one-dimensional case 
than for the three-dimensional anisotropic case considered in the main text.
The fact that
the build-up of ``wiggles'' is much slower in the one-dimensional case than in
the case considered in the main
text where there is a transfer of flux from the transverse to the axial degrees of freedom
underscores that collisions in three
spatial dimensions are much more ``effective'' 
than in one spatial dimension.
This conclusion is consistent with earlier
studies that worked with larger particle numbers and initial states 
that were closer to equilibrium~\cite{Monroe, Smith, DeMarco, jin, foot}.
We find similar behavior for other finite $g_{\text{1D}}$.
For infinitely large $g_{\text{1D}}$,
``wiggles'' are absent entirely.
This can be attributed to the scale invariance of the one-dimensional harmonically trapped two-atom system with diverging $g_{\text{1D}}$.
In contrast,
``wiggles'' develop for two atoms in an anisotropic trap with infinitely large $a_s$ due to the fact that the external confinement
is characterized by two or more harmonic oscillator lengths.

\end{document}